# Modelling emotional dynamics in Amrita Pritam's life: A nonlinear perspective on longing and companionship

Vinod Patidar[1,*] and Tanu Singh[2,*]

*[1,2]School of Computer Science, UPES, Dehradun, Uttarakhand, India*

[1]vinod.patidar@ddn.upes.ac.in; [1]patidar.vinod@gmail.com
[2]tanu.singh@ddn.upes.ac.in
*Corresponding Authors

**Abstract:**

Amrita Pritam (1919–2005) was one of the most prominent writers of modern Indian literature, particularly in Punjabi and Hindi. Her literary-biographical context provides a meaningful setting for exploring how memory, longing, companionship, and emotional regulation may be represented through simple dynamical models. This study develops three single-state ordinary differential equation models centred on two important relational references in her life: Sahir Ludhianvi, represented through memory, absence, recollection, and longing, and Imroz, represented through companionship, care, continuity, and support. The first one is a model that describes the Sahir-related longing through memory-driven excitation and natural emotional decay (fading); the second one models the Imroz-related companionship using nonlinear stabilization around a preferred emotional state; and the third one is a unified model that combines both Sahir's and Imroz's influences in a single state framework. We use analytical solutions, equilibrium conditions, local stability analysis, and numerical simulations to investigate the decay and persistence of emotional states, their convergence toward stable equilibria, and the effect of threshold conditions on system behaviour. The results show that memory-related reinforcement may sustain the modelled emotional state under suitable stimulus conditions. In contrast, Imroz-related companionship may drive the system toward a stable upper equilibrium, provided that the initial conditions lie within the admissible range. The unified model shows how memory-driven longing and companionship-driven stabilisation can jointly shape the same modelled emotional trajectory. The study is conceptual and mathematical rather than empirical or psychological. It does not attempt to measure Amrita Pritam's private emotions or estimate model parameters from biographical data. Instead, its contribution is to introduce a simple nonlinear modelling framework for exploring memory, decay (fading), companionship, equilibrium, and threshold effects within a literary-biographical context.

## 1. Introduction

Human emotions are dynamic in nature; they change with time. The emotions may fade, persist, revive, or stabilize depending on memory, absence, companionship, social circumstances, and creativity. Such time-dependent nature of emotions makes them suitable for dynamical modelling with an interest not in measuring emotions exactly, but rather in understanding the patterns of persistence, decay, regulation, and transition in emotional behaviour. Dynamical systems, specifically differential equations, can effectively model these changes and describe how emotional states shift under decay from absence, external stimuli, stabilisation through continuous support and companionship, and persistence through recollection. Such dynamic modelling allows us to understand qualitative changes in emotional states, such as fading memories, equilibrium, and transitions between growth and decay, as well as the conditions under which emotional states remain stable or bounded under competitive influences. The purpose of such modelling is not to quantify human emotions but to explain and understand the mechanisms by which emotional changes happen over time qualitatively.

The life of Amrita Pritam, a Punjabi-Hindi literary voice, offers an important context for such a dynamical exploration. Amrita Pritam's work comprises love, loss, partition, memory, self-reflection, and personal freedom, with the major emotional presence of Sahir Ludhianvi and Imroz that mimic the contrasting modes of emotional influence, which can be modelled and interpreted dynamically. The emotional presence of Sahir in Amrita's life brought in longing, memory, absence, poetic recollection, and unfulfilled attachment. Although Sahir's presence in her life was continuous, his influence is felt significantly through her literary imagination. In dynamical modelling terminology, it is an emotional stimulus that may be strong at a certain moment, persist through memory, and decay over time. Contrary to this, the emotional presence of Imroz in Amrita's life brings in the elements of companionship, care, continuity, acceptance, and creative support. Imroz's support, contrary to Sahir's intermittent and memory-induced influence, is steady and stabilizing, which may be modelled in terms of a companionship-based model regulated around a preferred state through linear and nonlinear stabilizing mechanisms (Pritam, 2001; Pritam, 2019; Trilok, 2006).

The mathematical modelling of human emotions and love dynamics is traditionally described in the works of Strogatz (1988), Rinaldi (Rinaldi, 1998a; Rinaldi, 1998b), Rinaldi and Gragnani (1998), Gottman et al. (2002), Sprott (2004), Rey (2010), and others, where differential equations, nonlinear response functions, delays, stability, and bifurcation analysis have been used to model and understand the romantic and sentimental dynamics. Nonlinear models have been used to study some of the legendary or literary love narratives, like Laura and Petrarch in the work of Rinaldi (1998b), and Layla and Majnun in the work of Sprott (2004), Jafari et al. (2016), and contemporary researchers. Much of this work is focused on the dyadic love interaction, marital dynamics, or mutual emotional response between the two individuals.

Our present work focuses on modelling the emotional state of a literary-biographical subject, Amrita Pritam, who is influenced by two distinct emotional sources: Sahir and Imroz. The first one sources memory, longing, and absence, and the other one sources care, companionship, and stabilization. The objective of our work is to develop and study the three single-state models that describe Amrita Pritam's emotional state. The first one is a model that describes the Sahir-related longing through memory-driven excitation and natural emotional decay (fading); the second one models the Imroz-related companionship using nonlinear stabilization around a preferred emotional state, and the third one is a unified model that combines both Sahir's and Imroz's influences in a single state framework.

The present study is conceptual mathematical work and does not claim to measure Amrita Pritam's emotions quantitatively nor to recreate the complete psychological complexity of her life. However, it provides a qualitative dynamical framework for memory-driven longing, companionship-driven stabilization, and their combined effect in the emotional life of Amrita Pritam.

**2. Related Work:**

Mathematical modelling of emotion, love, and attachment has been an established area of work, although comparatively small, within an intersection of nonlinear dynamics, applied mathematics, and social sciences. In such studies, emotions are treated as time-dependent variables, and their evolution is examined under the influence of factors such as attraction, memory, reaction, decay, feedback, stability, and instability. Our present work belongs to this class, but its purpose is not to present any general theory or dynamical representation of love or emotion dynamics; rather, it is focused on Amrita Pritam's literary-biographical context and its modelling as a dynamical system evolving with time, describing Amrita Pritam's emotional state qualitatively.

The earliest and most widely known mathematical modelling of romantic interaction between two individuals is credited to Strogatz (1988). In his work, coupled differential equations were used to represent the feelings of two individuals and how these feelings change under the influence of each other's response. This work showed that, although an idealized model, it can exhibit a wide variety of dynamical behaviour, interpreted as attraction, avoidance, oscillation, stability, and divergence. The most important contribution is that it showed love and emotions can be modelled and studied as dynamical systems.

Rinaldi (1998a) proposed a systematic linear framework incorporating self-response, response to a partner, and appeal. This framework provided a useful baseline for distinguishing between one-variable and nonlinear models. Since linear systems exhibit relatively limited long-term dynamics, the framework also highlights the need to introduce additional state variables or nonlinear terms when more complex behaviour is being examined.

Rinaldi and Gragnani (1998) developed this area further by studying the love dynamics between secure individuals, incorporating appeal, response behaviour, relationship stability, and distinguishing between fragile and robust relationships. This work is relevant to the present study because it demonstrates how emotional dynamics can be examined qualitatively through the analysis of equilibrium points and their stability. It also illustrates the potential of nonlinear response functions to model variations in emotional states that may not be adequately described by linear models.

Gragnani et al. (1997) analyzed two-person relationship models by classifying them according to attachment style and synergism. They showed that secure or non-synergistic couples exhibit only stationary behaviour, whereas cyclic emotional dynamics arise when insecurity and synergism coexist within a relationship. Their local stability and bifurcation analyses identified how responsiveness and synergism shape relational dynamics and further showed that ageing may have a stabilising effect on emotional states, consistent with common psychological observations. This study provides a useful foundation for the present work, which examines not only convergence to equilibrium but also the emergence of sustained oscillatory patterns in relationship dynamics.

Gottman et al. (2002) developed a framework for understanding emotional and relational processes by modelling marital interactions using nonlinear terms. Unlike purely illustrative models of romantic relationships, their work adopts an empirical approach, connecting patterns of interaction to relationship stability. Our present work differs from this work in the sense that it is not based on measured interaction data; rather, it is a conceptual and literary-biographical model. However, the ideas of feedback, stability and emotional regulation remain relevant. Rey (2010) presented a model for sentimental dynamics that incorporated long-term relationship behaviour and marital dissolution. His work is relevant to the present work, as it shows how decay, persistence, threshold-like behaviour and long-term stability can be studied within a dynamical framework. His work focused on prediction of marital dissolution, which our model does not consider; however, our model represents the memory-driven longing and companionship-driven stabilization as a single emotional state.

Sprott (2004) modelled the love dynamics by considering nonlinearities, love triangles, and more complex behaviour. His work emphasized that simple emotional models can successfully describe the rich and complex behaviours when nonlinear terms and additional relational influences are introduced. Our work also proposes to include nonlinear regulation in the Amrita-Imroz and unified model and focuses on decay, boundedness, equilibrium, and thresholds and ignores complex chaotic behaviours. Dercole and Rinaldi (2014) provided a useful framework for analysing literary or cinematic relationships as nonlinear systems. The study demonstrates how a narrative involving three individuals can be represented through a dynamical model exhibiting sensitive and potentially unpredictable behaviour. It also supports the use of cultural biography as a context for modelling, although it should not be regarded as empirical validation of psychological claims.

The work of Rinaldi (1998b) on Laura and Petrarch is especially relevant here, as it is one amongst the literary applications of love dynamics. It showed how historical literary emotional relationships can be analyzed through nonlinear dynamic modelling. This work is highly relevant to the present work, as Amrita Pritam’s emotional state is described through literary, autobiographical, and biographical material rather than through experimental measurements. Therefore, Rinaldi’s work is an important precedent for the present work as the modelling of literary emotional relationships. Although the present study differs in its structure from Rinaldi’s work, as it models the emotional state of one literary subject under two contrasting influences rather than modelling mutual romantic interaction between two lovers. Jafari et al. (2016) modelled a culturally known love narrative of Layla and Majnun, which is another relevant example. It supports the broader possibility of applying mathematical models to literary/legendary emotional narratives. The present study conceptualizes Amrita Pritam’s emotional state as a single-subject system influenced by two contrasting relational sources: Sahir, associated with memory and longing, and Imroz, associated with companionship and stability.

Delay and memory (Liao and Ran, 2007; Matsumoto, 2017; Tabatabaei et al., 2014) are important themes in emotional dynamics, and studies on delayed love-affair models showed that memory, forgetting, and response lag significantly change the long-term behaviour. It is also important in the present context of Amrita Pritam, as the Sahir-related component is associated with memory and recollection. However, the present manuscript keeps the modelling simple by avoiding the delay and representing the Sahir-related influence through external stimuli such as constant and exponentially decaying input. Delay-based or memory-kernel-based models may be considered in later studies.

Although the existing literature provides a solid foundation for modelling emotions and love, most available models focus on mutual interactions, marital stability or dissolution, love triangles, or literary/legendary dyads. These studies primarily address how the emotional state of one person influences that of another and how their interaction evolves over time. Although such models provide valuable insights, they do not directly address the specific situation examined in the present manuscript.

Our study focuses on a single literary-biographical subject under the influence of two qualitatively different sources. Sahir Ludhianvi represents longing, memory, absence, and poetic recollection, whereas Imroz represents companionship, care, continuity, and emotional stabilization. This distinction motivates us to propose and analyze three models: a Sahir-related memory-decay model, an Imroz-related companionship-stabilization model, and a unified model that combines both influences.

The novelty of the present work does not lie in proposing a new mathematical model of emotion and love; rather, it lies in applying a single-state nonlinear dynamical modelling framework to the literary-biographical emotional life of Amrita Pritam. Thus, the present study offers a first mathematical abstraction of Amrita Pritam's emotional life by representing it through a single state variable, while leaving multivariate modelling and the incorporation of additional emotional dimensions for future work.

## 3. Biographical and Conceptual Background

### *Amrita Pritam: Literary and Emotional Context*

Amrita Pritam (1919-2005) was one of the major literary voices in modern Indian literature who wrote primarily in Punjabi and Hindi on the themes of love, loss, memory, partition, personal freedom, womanhood, and emotional self-reflection. Her literary imagination was deeply influenced by the partition of India, while her emotional relationship remains important in personal writings and public memory (Pritam, 2001; Pritam, 2019; Nandrajog and Srivastava, 2023).

This section does not aim to provide the complete biography of Amrita Pritam; it identifies and presents the selected emotional patterns from her literary-biographical context that are relevant and subject to modelling. Sahir Ludhianvi and Imroz are important, as they represent two contrasting forms of emotional presence in her life. Sahir is mainly associated with longing, distance, memory, and poetic recollection, whereas Imroz is associated with companionship, care, continuity, and support.

### *Sahir Ludhianvi: Longing and Memory*

The Amrita-Sahir association is often described as an intense but largely unfulfilled emotional attachment that is remembered generally through silence, separation, literary sensitivity, and persistence of feeling despite limited companionship. Sahir's importance in this context lies not only in direct interaction but also in his continued presence through memory and imagination (Pritam, 2001; Pritam, 2019).

Therefore, in this study, Sahir is understood as a figure of longing and remembrance; his influence is not treated as a steady domestic presence; rather, it is treated as an emotional force linked through absence, recollection, and inward recurrence. This interpretation develops the background of the first model presented in Section 4, where the Sahir-related component is included in terms of memory-driven persistence and natural fading.

### *Imroz: Companionship and Stability*

In Amrita Pritam's life, Imroz brings in a different emotional and relational dimension that is associated with companionship, acceptance, care, and creative support. Unlike Sahir's association, which is responsible for longing and absence, Imroz is linked with continuity and presence that is important for understanding the stabilizing aspects of the later emotional and creative life of Amrita Pritam (Trilok, 2006; Pritam, 2019)

In this study, Imroz is therefore treated as a source of companionship and stability. His presence includes emotional modes based on support rather than distance, continuity rather than interruptions, and care rather than unresolved longing. This provides the foundation of our second model presented in Section 5, where companionship is considered in relation to emotional regulation and stability.

***Conceptual Transition to Modelling***

The contrast between Sahir and Imroz provides the conceptual foundation for the three models developed in the following sections. Sahir-related influence is considered in relation to longing, memory, and gradual fading, while Imroz-related influence is considered in relation to companionship, care, and stabilization. The first model in Section 4 studies the Sahir-related emotional response, the second model in Section 5 studies the Imroz-related stabilizing response, and the third one in Section 6 combines both influences within a single-state dynamical formulation.

## 4. Model I: Amrita-Sahir Memory-Driven Emotional Model

***Model Formulation***

The first model describes the Sahir-related emotional component in Amrita Pritam's literary-biographical context. As discussed in the previous section, Sahir Ludhianvi is interpreted here as a memory-oriented influence associated with longing, absence, recollection, and poetic association. To represent this mathematically, a single emotional-state variable $A(t)$ is introduced, where $A(t)$ denotes Amrita's modelled emotional state at time $t$.

The Sahir-related emotional stimulus is denoted by $S(t)$. This function represents the effect of memory, poetic recollection, symbolic presence, emotional recall, or other inward triggers associated with Sahir. The evolution of $A(t)$ is described by the first-order differential equation.

$$\frac{dA}{dt} = \alpha S(t) - \beta A(t),$$

where $\alpha > 0$ represents the sensitivity/responsiveness of the emotional state to the Sahir-related stimulus, and $\beta > 0$ denotes the natural decay (fading) rate of emotional intensity.

The term $\alpha S(t)$ here represents the contribution of Sahir-related recollection to the modelled emotional state. When memory, poetic association, or inward recall is active, this term tends to maintain or increase $A(t)$. However, the term $-\beta A(t)$ represents the gradual fading of the same state when such reinforcement is absent. Therefore, the model describes Amrita's emotional intensity as being influenced by two competing effects: Sahir-related recollection sustains or reinforces it, while natural decay over time gradually weakens it.

This model only provides a simple mathematical form for qualitatively understanding how a memory-related emotional response may persist, decline, or approach a stable level under different assumptions about $S(t)$.

***Analytical Solution***

The governing equation can be written as

$$\frac{dA}{dt} + \beta A(t) = \alpha S(t).$$

Multiplying by the integrating factor $e^{\beta t}$, we obtain

$$\frac{d}{dt}\left(e^{\beta t} A(t)\right) = \alpha e^{\beta t} S(t).$$

Assuming $A(0) = A_i$ and integrating with respect to time

$$e^{\beta t} A(t) - A_i = \alpha \int_0^t e^{\beta \tau}\, S(\tau)\, d\tau.$$

Therefore,

$$A(t) = e^{-\beta t}\left[A_i + \alpha \int_0^t e^{\beta \tau}\, S(\tau)\, d\tau\right].$$

The first part of the solution on the right-hand side, i.e.,$A_i e^{-\beta t}$, represents the fading effect of the initial emotional state. While the integral term on the right side

$$\alpha e^{-\beta t} \int_0^t e^{\beta \tau}\, S(\tau)\, d\tau,$$

represents the cumulative effect of Sahir-related stimuli over time. It is weighted by an exponential weight factor; therefore, the emotional state $A(t)$, depends not only on the current stimulus but also on previous stimuli, whose effects gradually decay over time. Hence, the initial emotional state $A_i$ does not disappear immediately; rather, its influence gradually decreases through the exponential decay term. At the same time, Sahir-related stimuli represented by $S(t)$, such as memory, poetic recollection, symbolic presence, or inward emotional recall, continue to contribute to the evolution of $A(t)$. Thus, the solution shows how the modelled emotional state is shaped jointly by natural fading and memory-driven reinforcement.

***Special Stimulus Cases***

Three simple forms of $S(t)$ are considered to illustrate different possible modes of Sahir related influence.

*Case I: Absence of Stimulus:*

If there is no continuing Sahir-related reinforcement, then $S(t) = 0$. The model reduces to $\dot{A} = -\beta A(t)$, with a solution $A(t) = A_i e^{-\beta t}$. This case represents pure exponential decay, i.e, it corresponds to a situation in which no memory, recollection, poetic association, or symbolic stimulus continues to reinforce the emotional state. The larger the value of $\beta$, the faster the decay (fading) and a smaller $\beta$ indicates slower fading and greater persistence.

*Case II: Constant Stimulus:*

If the Sahir-related stimulus remains constant, then $S(t) = S_0$, where $S_0 > 0$. The model becomes $\dot{A} = \alpha S_0 - \beta A(t)$, and the solution is

$$A(t) = A_i e^{-\beta t} + \frac{\alpha S_0}{\beta}\left(1 - e^{-\beta t}\right).$$

As $t \to \infty$, the solution approaches the equilibrium value

$$A^* = \frac{\alpha S_0}{\beta}.$$

Thus, when memory or recollection remains continuously active, the emotional state does not decay to zero. Instead, it approaches a stable level determined by the strength of the stimulus and the rate of natural fading.

*Case III: Exponentially Decaying Stimulus:*

A more realistic possibility is that the Sahir-related stimulus is initially strong but gradually weakens with time. This can be modelled by $S(t) = S_0 e^{-\lambda t}$, where $S_0 > 0$ is the initial stimulus and $\lambda > 0$ is the decay rate of the stimulus. Substitution into the model gives $\dot{A} = \alpha S_0 e^{-\lambda t} - \beta A(t)$. For $\beta \neq \lambda$, the solution is

$$A(t) = A_i e^{-\beta t} + \frac{\alpha S_0}{\beta - \lambda}\left(e^{-\lambda t} - e^{-\beta t}\right).$$

In the special case $\beta = \lambda$, the solution becomes $A(t) = A_i e^{-\beta t} + \alpha S_0 t e^{-\beta t}$. For the exponentially decaying stimulus, the Sahir-related influence is strongest at the initial stage and then decreases with time. This case is useful because it avoids two extreme situations: (i) the stimulus disappears completely, or (ii) it remains constant forever.

***Numerical Simulation and Interpretation***

To examine the model's behaviour under different conditions, three forms of stimulus are considered: no stimulus, a constant stimulus, and an exponentially decaying stimulus. For the numerical simulations, the initial value is set to $A_i = 5$, representing a pre-existing level of modelled emotional intensity at the starting time. The parameter values $S_0 = 10$, $\alpha = 1.2$, $\beta = 0.4$, and $\lambda = 0.1$ are selected to generate clearly distinguishable qualitative trajectories for the three stimulus cases. These values are illustrative rather than empirical; they are used to demonstrate pure decay, convergence to a sustained equilibrium, and gradual attenuation under a decaying stimulus. Figure 1 shows the three forms of the Sahir-related stimulus $S(t)$. In the first case, $S(t) = 0$, no reinforcing influence is present. In the second case, $S(t) = S_0$, the stimulus remains constant. In the third case, $S(t) = S_0 e^{-\lambda t}$, the stimulus begins at $S_0$ and gradually decreases toward zero. Figure 2 shows the corresponding Amrita's emotional state $A(t)$. In the absence of a stimulus, the emotional state decreases monotonically toward zero. Under a constant stimulus, the emotional state approaches a positive equilibrium. Under an exponentially decaying stimulus, the response reflects a transient, memory-driven effect: the emotional state may remain higher for some time but eventually declines as the stimulus weakens.

The numerical results illustrate three qualitatively different emotional patterns. The first is pure fading, produced by natural decay in the absence of reinforcement. The second is sustained persistence, produced by continuous reinforcement, under which the emotional state approaches a stable positive equilibrium. The third is gradual attenuation, produced by a stimulus that itself fades with time. The Amrita-Sahir model provides a minimal mathematical representation of memory-driven emotional persistence. Its main purpose is to show how an emotional state associated with longing and recollection may evolve under different assumptions about the reinforcing stimulus without assigning empirical or psychological meaning to the selected parameter values.

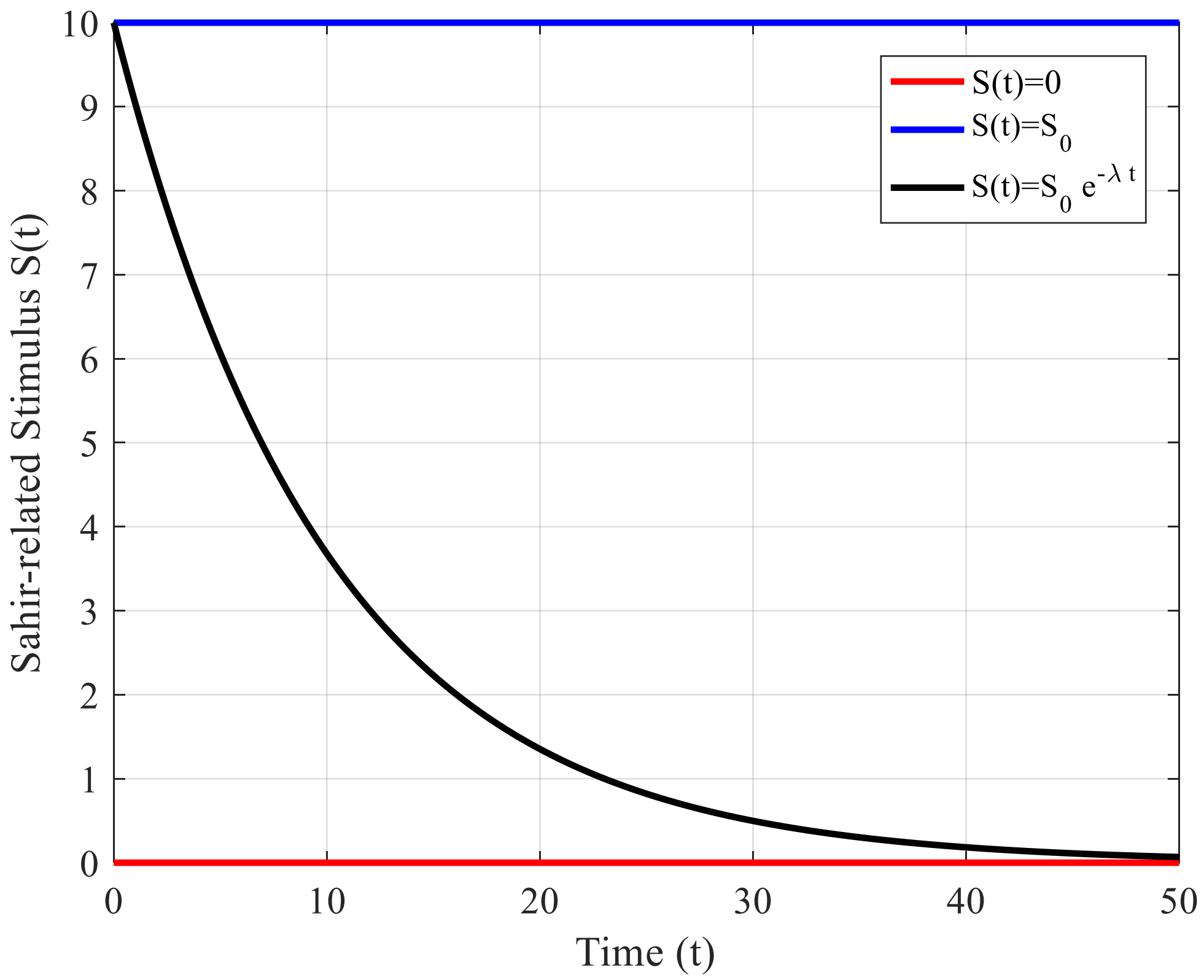

**Figure 1.** Sahir-related stimulus functions $S(t)$ under three assumptions: absence of stimulus, constant stimulus, and exponentially decaying stimulus.

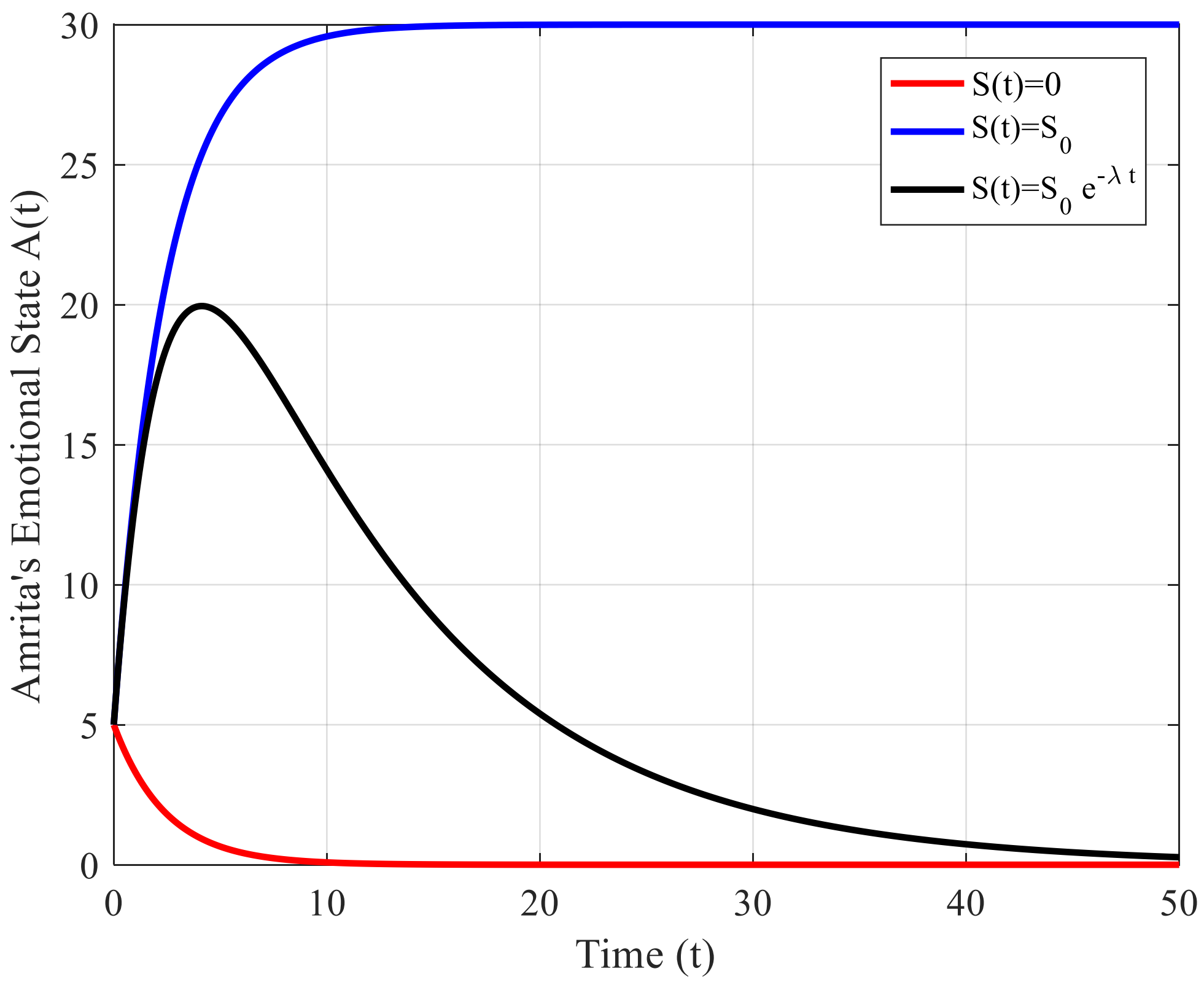

**Figure 2.** Amrita's emotional state $A(t)$ in the Amrita-Sahir model for different Sahir-related stimulus functions $S(t)$.

In the literary-biographical interpretation considered here, the Amrita–Sahir association is represented in the model as a memory-based influence rather than as a continuing form of companionship. The three stimulus functions correspond to different idealized modes of this influence: complete fading when memory-related stimuli (such as recollection, poetic memories, symbolic presence, and inward emotional recall) provide no further reinforcement; sustained emotional intensity when such stimuli remain continuously active; and gradual attenuation when the memory-related stimuli progressively weaken over time. Within the conceptual limits of the model, these cases illustrate how longing and recollection may interact with natural emotional decay to produce fading, persistence, or temporary emotional elevation. The model is not intended to measure Amrita Pritam's private emotions; it offers only a mathematical abstraction inspired by the literary-biographical association between Amrita Pritam and Sahir Ludhianvi.

The next section considers a contrasting emotional mode associated with Imroz, where the emphasis shifts from memory-driven longing to companionship-driven stabilization.

## 5. Model II: Amrita-Imroz Companionship-Stabilization Model

### *Model Formulation*

The second model describes the Imroz-related component in Amrita Pritam's literary-biographical context. In contrast to the Sahir-related model, where the emphasis was on memory, longing, and natural decay, the present model is concerned with companionship, care, continuity, and emotional stabilization. Imroz is therefore interpreted as a steady supportive influence rather than a fading memory-based stimulus.

Let $A(t)$ denote Amrita's modelled emotional state at time $t$, and let $I(t)$ represent the Imroz-related influence. Here, $I(t)$ may be interpreted as an input associated with companionship, care, continuity, acceptance, and creative support. The proposed model is

$$\frac{dA}{dt} = \gamma I(t) - \delta_1(A - A_0) - \delta_2(A - A_0)^2,$$

where $\gamma > 0$ measures the strength of the Imroz-related supportive contribution, $\delta_1 > 0$ is the linear regulation coefficient, $\delta_2 > 0$ is the nonlinear regulation parameter, and $A_0$ denotes a reference or preferred emotional level. The term $\gamma I(t)$ represents the contribution of Imroz-related companionship, and the term $-\delta_1(A - A_0)$ describes linear regulation around the reference level $A_0$. When $A(t) > A_0$, this term acts to reduce $A(t)$, while for $A(t) < A_0$, it acts to increase $A(t)$. The nonlinear term $-\delta_2(A - A_0)^2$ introduces a threshold-type effect. It strengthens the downward correction when $A(t)$ is above $A_0$, but it should not be interpreted as a globally restoring term for all possible values of $A(t)$.

This point is important for the interpretation of the model. The quadratic term helps produce a stable upper equilibrium along with a lower unstable threshold. Therefore, the model should be interpreted within a meaningful emotional range and for admissible initial conditions. Its purpose is not to provide a complete psychological law, but rather to represent how companionship and emotional regulation may lead to a stable, modelled emotional state.

### *Equilibrium Analysis*

To study the long-term behaviour of the model, consider the case of constant Imroz-related influence, $I(t) = I_0$, where $I_0 > 0$. The governing equation becomes

$$\dot{A} = \gamma I_0 - \delta_1(A - A_0) - \delta_2(A - A_0)^2.$$

At equilibrium, $\gamma I_0 - \delta_1(A - A_0) - \delta_2(A - A_0)^2 = 0$. Considering $x = A - A_0$, the equilibrium condition becomes $\delta_2 x^2 + \delta_1 x - \gamma I_0 = 0$, that further gives

$$x_{1,2} = \frac{-\delta_1 \pm \sqrt{\delta_1^2 + 4\delta_2\gamma I_0}}{2\delta_2}.$$

Therefore, the equilibrium states are

$$A_{1,2}^* = A_0 + \frac{-\delta_1 \pm \sqrt{\delta_1^2 + 4\delta_2\gamma I_0}}{2\delta_2}.$$

Since $\delta_1^2 + 4\delta_2\gamma I_0 > 0$, the model has two real equilibrium points for positive parameter values. The root with the positive sign gives the upper equilibrium, while the root with the negative sign gives the lower equilibrium.

The upper equilibrium lies above $A_0$, since

$$\frac{-\delta_1 + \sqrt{\delta_1^2 + 4\delta_2\gamma I_0}}{2\delta_2} > 0,$$

and the lower equilibrium lies below $A_0$. Thus, the model naturally produces two equilibrium levels: one stable level associated with companionship-supported regulation, and one lower threshold separating different types of motion.

***Stability Analysis***

Define $f(A) = \gamma I_0 - \delta_1(A - A_0) - \delta_2(A - A_0)^2$. Then the model can be written as $\dot{A} = f(A)$ and the derivative of $f(A)$ is $f'(A) = -\delta_1 - 2\delta_2(A - A_0)$. At an equilibrium $A^*$, the sign of $f'(A^*)$ determines local stability. If $f'(A^*) < 0$, then the equilibrium is locally asymptotically stable. If $f'(A^*) > 0$, then the equilibrium is unstable. Using $x = A - A_0$, we have $f'(A^*) = -\delta_1 - 2\delta_2 x^*$. For the upper equilibrium,

$$x^* = \frac{-\delta_1 + \sqrt{\delta_1^2 + 4\delta_2\gamma I_0}}{2\delta_2}.$$

Substitution gives

$$f'(A^*) = -\sqrt{\delta_1^2 + 4\delta_2\gamma I_0} < 0.$$

Therefore, the upper equilibrium is stable.

For the lower equilibrium,

$$x^* = \frac{-\delta_1 - \sqrt{\delta_1^2 + 4\delta_2\gamma I_0}}{2\delta_2}.$$

Substitution gives

$$f'(A^*) = \sqrt{\delta_1^2 + 4\delta_2\gamma I_0} > 0.$$

Therefore, the lower equilibrium is unstable.

The analysis indicates that the model exhibits a threshold-dependent structure. For all the initial conditions above the lower equilibrium, the solution approaches the upper, stable equilibrium and all initial conditions below the lower equilibrium do not fall within the stabilizing regime described by the model. Consequently, the formulation should not be considered as globally bounded for all possible initial conditions. Rather, it represents a companionship-based stabilization mechanism that is valid within a specified admissible range of the modelled emotional state.

***Numerical Simulations and Interpretation***

Numerical simulations are used to illustrate the behaviour of the Amrita-Imroz model under different assumption for the Imroz-related input function $I(t)$. We have considered the following three representative cases :(i) $I(t) = 0$ i.e. the absence of Imroz-related input, (ii) $I(t) = I_0$ i.e., the steady companionship or continuing support, and (iii) $I(t) = I_0 e^{-\mu t}$. i.e., a gradually weakening supportive influence. These cases are not intended as empirical reconstructions, but as simple mathematical situations through which the qualitative behaviour of the model can be observed. For numerical illustration, the following parameter values are used: $A(0) = A_i = 3$, $A_0 = 5$, $I_0 = 4$, $\gamma = 1$, $\delta_1 = 0.5$, $\delta_2 = 0.1$, $\mu = 0.05$. The values are selected to observe the effect of companionship-related input, linear regulation, and nonlinear threshold behaviour. They are illustrative values, not estimated biographical parameters.

Figure 3 shows the three forms of the Imroz-related input $I(t)$. When $I(t) = 0$, no external companionship-related influence is present. When $I(t) = I_0$, the influence remains constant and represents sustained support. When $I(t) = I_0 e^{-\mu t}$, the influence decreases gradually with time. Figure 4 presents the response of the corresponding linear model, obtained by putting $\delta_2 = 0$. Under this condition, the emotional intensity $A(t)$ is governed by $I(t)$ and the linear regulation term. The solutions therefore move gradually toward the level determined by $A_0$ or the chosen input $I(t)$. Since the quadratic term is removed, the lower-threshold behaviour seen in the nonlinear model does not appear. Figure 5 shows the Amrita's emotional intensity $A(t)$ in the Amrita-Imroz model by considering both linear and nonlinear terms. For $I(t) = I_0$, the trajectory moves toward the stable upper equilibrium obtained in the preceding analysis. For $I(t) = I_0 e^{-\mu t}$, the input is initially active and then weakens, forcing the emotional intensity to rise initially before settling under the effect of the regulation terms. When $I(t) = 0$, the motion is governed only by the terms involving $A - A_0$. This interpretation is valid for all initial values chosen above the lower unstable equilibrium. Figure 6 presents the phase-line representation for the case of a constant Imroz-related input i.e., $I(t) = I_0$. The zeros of $f(A)$, where it crosses the horizontal axis, correspond to the equilibrium values of the Amrita-Imroz model. The sign of $f(A)$ on either side of these points indicates the direction of motion. From this diagram, the lower equilibrium is observed as a threshold value, whereas the upper equilibrium is stable and attracts nearby trajectories.

In the literary-biographical interpretation used in this study, the Amrita–Imroz model treats companionship as a continuing source of emotional support rather than a temporary stimulus. This influence includes care, acceptance, continuity, and companionship. The upper stable equilibrium may be interpreted metaphorically as a persistent emotional state maintained through continued companionship and regulatory effects. By contrast, the lower unstable equilibrium denotes a threshold below which the model's stabilizing mechanism ceases to maintain that state. The model therefore provides a mathematical abstraction of how enduring companionship may counteract emotional fluctuations and promote persistence and stabilization over time. This interpretation provides a qualitative mathematical representation inspired by the literary-biographical relationship between Amrita and Imroz.

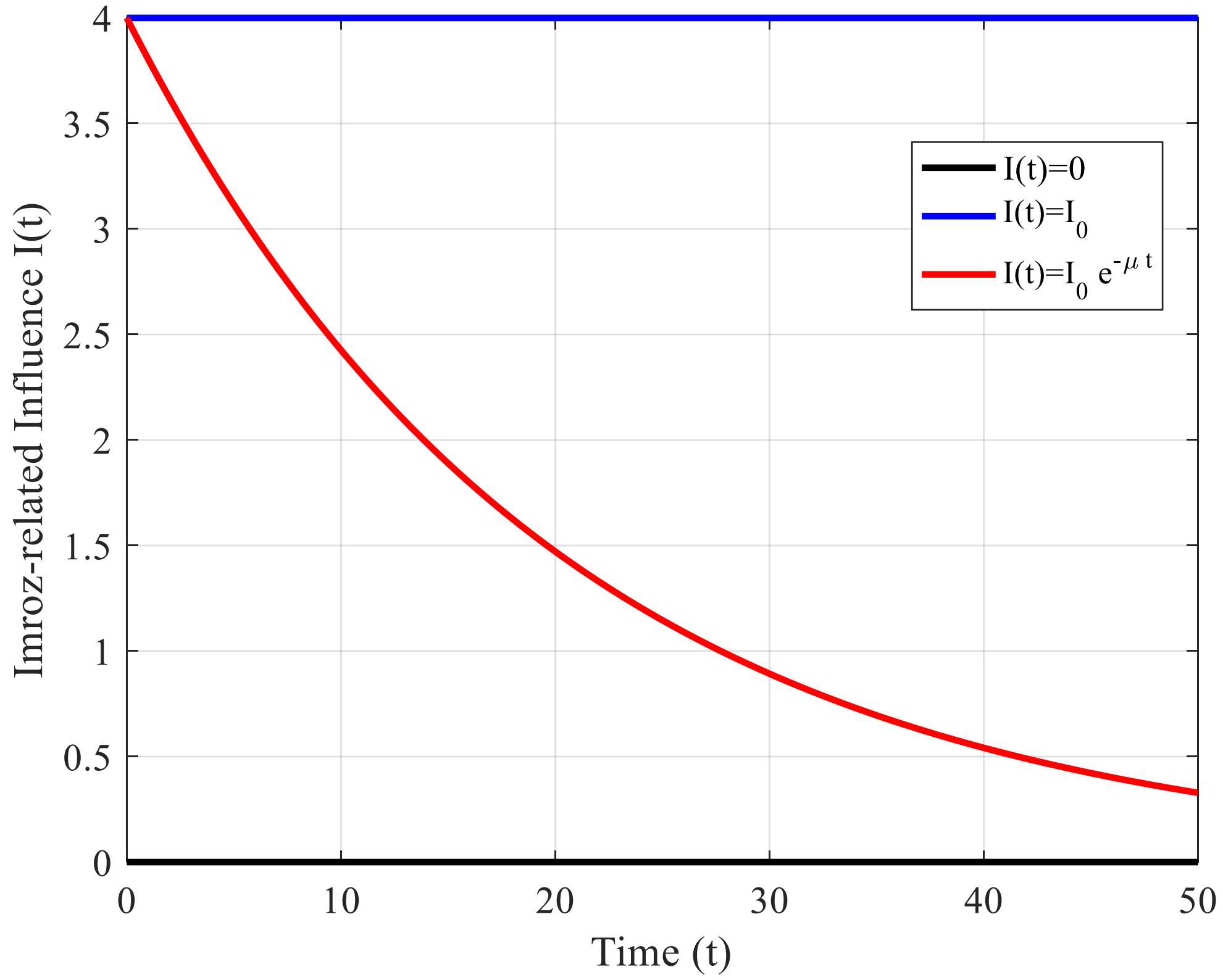

**Figure 3.** Imroz-related input functions $I(t)$ under three cases: absence of influence, constant influence, and exponentially decaying influence.

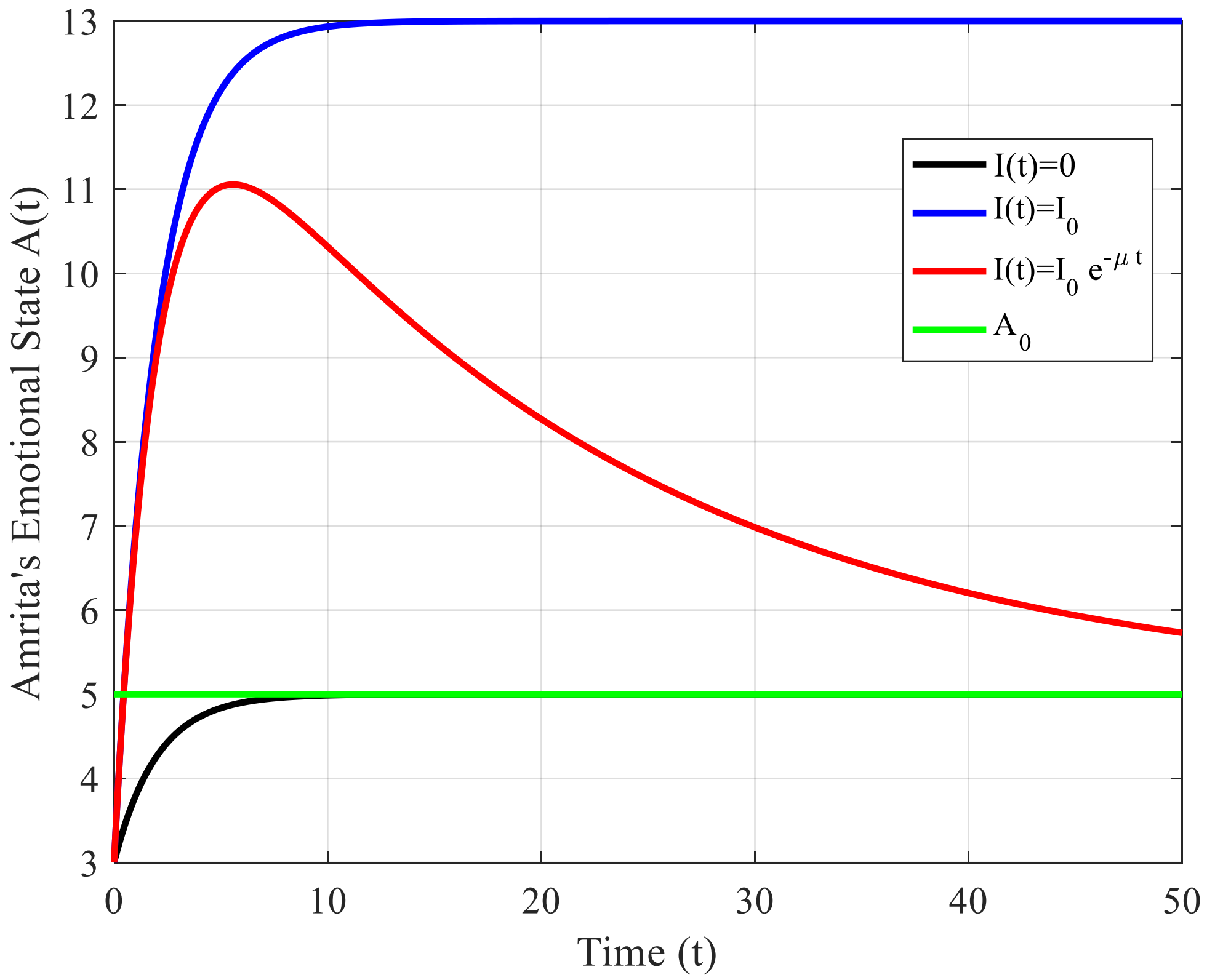

**Figure 4.** Amrita's Emotional state $A(t)$ in the linear Amrita-Imroz model for different Imroz-related input functions $I(t)$

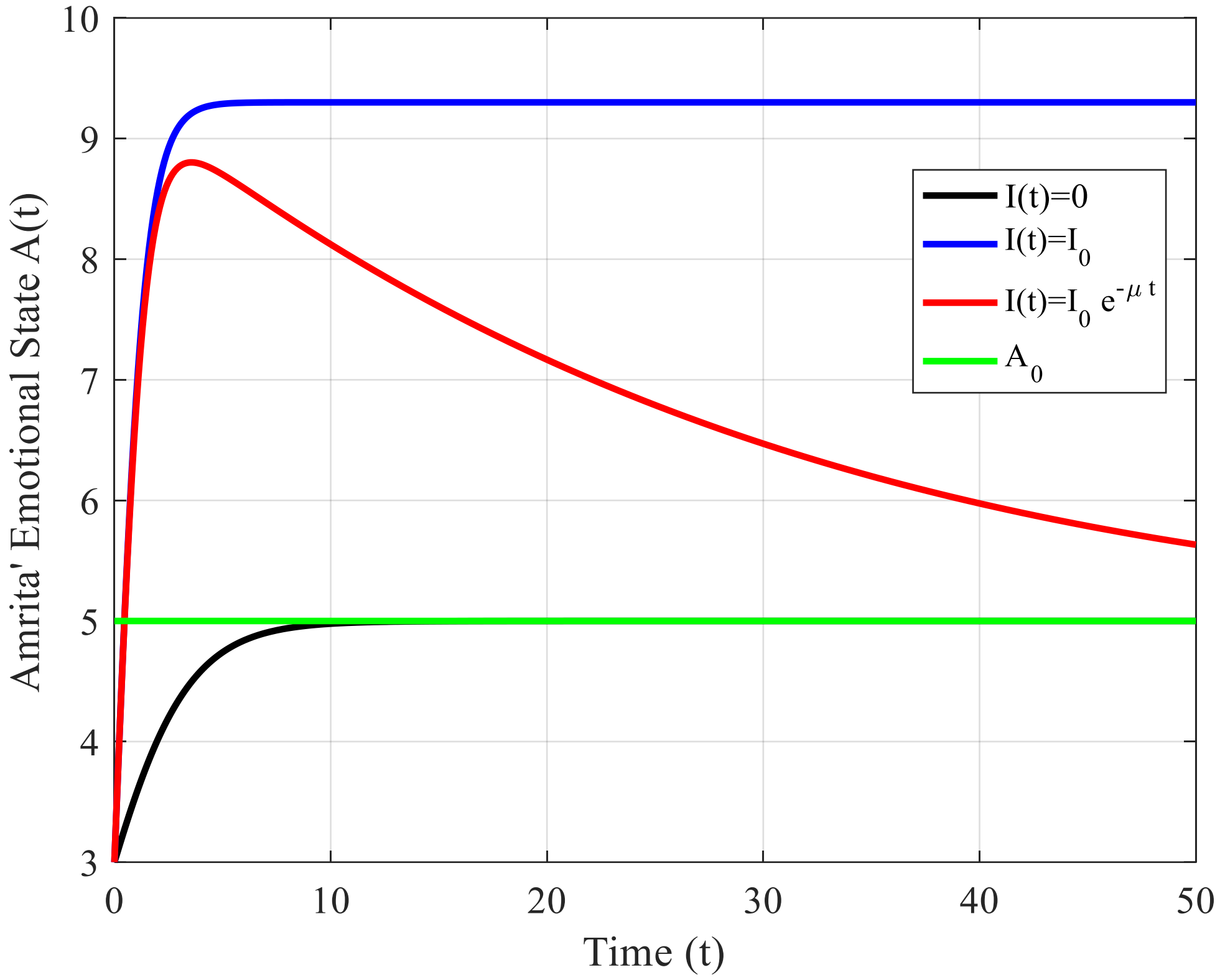

**Figure 5.** Amrita's Emotional state $A(t)$ in the nonlinear Amrita-Imroz model for different Imroz-related input functions $I(t)$

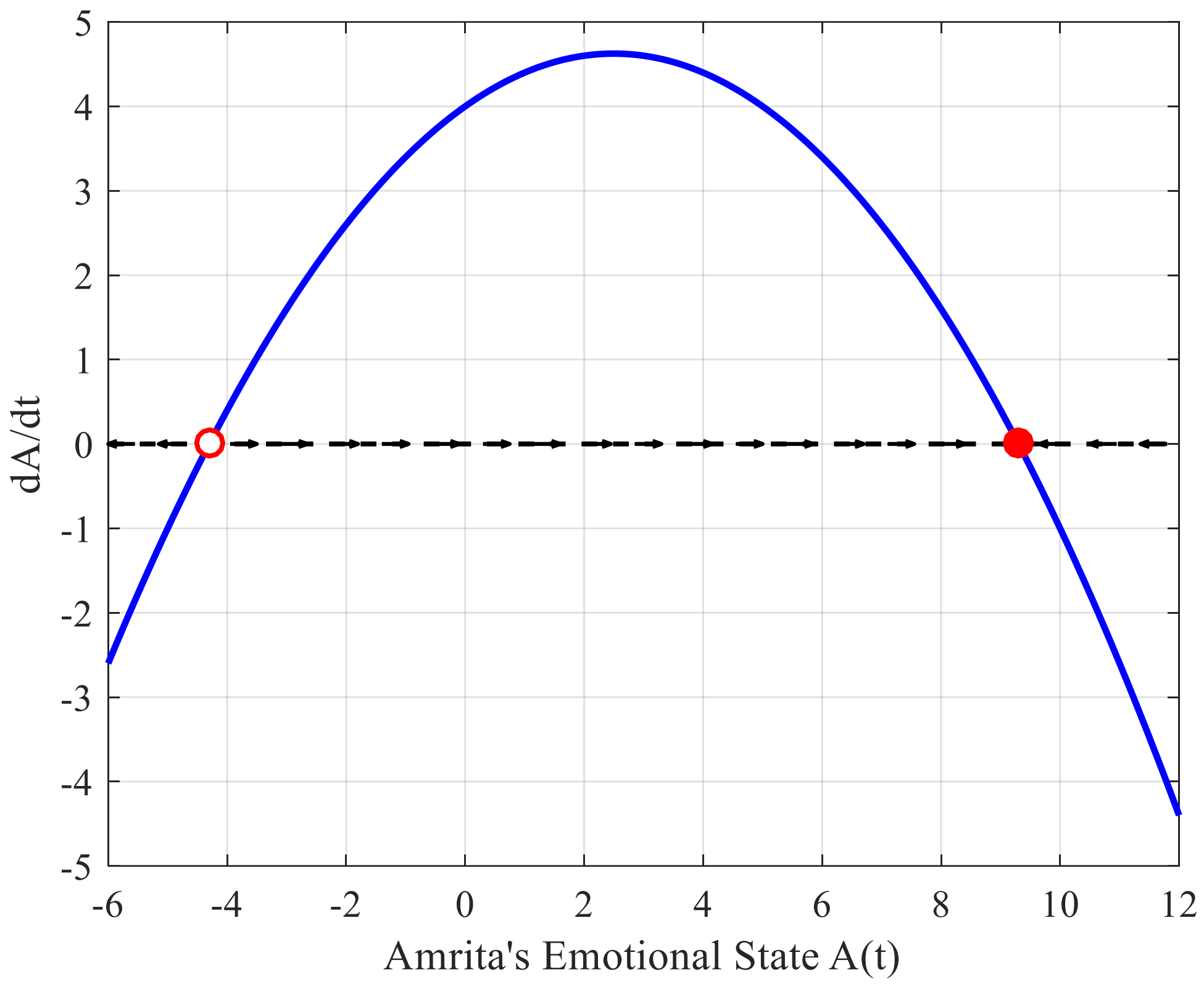

**Figure 6.** Phase-line diagram of the Amrita-Imroz model considering constant Imroz-related input, i.e., $I(t) = I_0$.

The next section combines the Sahir-related stimulus and the Imroz-related input in a single mathematical framework.

## 6. Model III: Unified Amrita-Sahir-Imroz Model

### 6.1 Unified Model Formulation

The first two models considered the Sahir-related and Imroz-related influences separately. Model I described the Sahir-related component through memory-driven reinforcement and natural decay. Model II described the Imroz-related component through companionship-related input and regulation around a reference level. The purpose of the present section is to place these two influences together in a single model equation.

Let $A(t)$, $S(t)$ and $I(t)$ respectively denote the modelled Amrita's emotional state, the Sahir-related stimulus and the Imroz-related input. Then the unified model is written as

$$\frac{dA}{dt} = \alpha S(t) - \beta A(t) + \gamma I(t) - \delta_1(A - A_0) - \delta_2(A - A_0)^2.$$

Here, $\alpha S(t)$ represents the contribution of Sahir-related memory, recollection, or symbolic presence. The term $-\beta A(t)$ represents natural emotional decay. The term $\gamma I(t)$ represents the contribution of Imroz-related companionship and support. The remaining terms, $-\delta_1(A - A_0)$ and $-\delta_2(A - A_0)^2$, represent regulation with respect to the reference level $A_0$.

Thus, the unified model combines two different mechanisms. The Sahir-related term introduces memory-based reinforcement, while the Imroz-related terms introduce support and regulation. The model remains a single-state formulation, but it allows the same state variable $A(t)$ to be affected by both memory-driven and companionship-driven components.

### 6.2 Reduction to Earlier Models

The unified model contains the earlier models as limiting cases, with one notational caution. If the Imroz-related input and regulation terms are removed, that is $I(t) = 0$, $\delta_1 = 0$, $\delta_2 = 0$, then the unified equation becomes $\dot{A} = \alpha S(t) - \beta A(t)$. This is exactly the Amrita-Sahir model developed in Section 4.

If the Sahir-related stimulus is removed, that is, $S(t) = 0$, then the equation becomes $\dot{A} = -\beta A(t) + \gamma I(t) - \delta_1(A - A_0) - \delta_2(A - A_0)^2$.
This is an Imroz-type stabilization model with an additional natural decay term $-\beta A(t)$. If one also sets $\beta = 0$, the equation reduces exactly to the Amrita-Imroz model of Section 5: $\dot{A} = \gamma I(t) - \delta_1(A - A_0) - \delta_2(A - A_0)^2$.

When both inputs are present i.e., $S(t) \neq 0$, $I(t) \neq 0$, the model describes the combined action of Sahir-related memory and Imroz-related companionship on the same emotional state. In this case, the model does not treat the two influences as mutually exclusive. Instead, it studies how both may enter the evolution of $A(t)$ through different mathematical terms.

### 6.3 Equilibrium Analysis Under Constant Inputs

For equilibrium analysis, consider constant inputs $S(t) = S_0$, $I(t) = I_0$, where $S_0 > 0$, $I_0 > 0$. The unified model becomes $\dot{A} = \alpha S_0 - \beta A + \gamma I_0 - \delta_1(A - A_0) - \delta_2(A - A_0)^2$.

At equilibrium, $\alpha S_0 - \beta A + \gamma I_0 - \delta_1(A - A_0) - \delta_2(A - A_0)^2 = 0$. Considering $x = A - A_0$, i.e, $A = x + A_0$ gives $\delta_2 x^2 + (\beta + \delta_1)x + \beta A_0 - \alpha S_0 - \gamma I_0 = 0$. Hence,

$$x_{\pm} = \frac{-(\beta + \delta_1) \pm \sqrt{(\beta + \delta_1)^2 + 4\delta_2(\alpha S_0 + \gamma I_0 - \beta A_0)}}{2\delta_2}.$$

Thus, the equilibrium states are

$$A_{\pm}^* = A_0 + \frac{-(\beta + \delta_1) \pm \sqrt{(\beta + \delta_1)^2 + 4\delta_2(\alpha S_0 + \gamma I_0 - \beta A_0)}}{2\delta_2}.$$

Real equilibria exist when the discriminant $\Delta = (\beta + \delta_1)^2 + 4\delta_2(\alpha S_0 + \gamma I_0 - \beta A_0) \geq 0$. When $\Delta > 0$, there are two equilibrium branches. Similar to the Amrita-Imroz model, the upper branch corresponds to $A_+^*$ that gives the stable level toward which admissible trajectories may move, and the lower branch corresponds to $A_-^*$ that describes the threshold.

**6.4 Stability Analysis**

Let $f(A) = \alpha S_0 - \beta A + \gamma I_0 - \delta_1(A - A_0) - \delta_2(A - A_0)^2$. Then the autonomous model can be written as $\dot{A} = f(A)$.
The derivative of $f(A)$ is $f'(A) = -\beta - \delta_1 - 2\delta_2(A - A_0)$.
At equilibrium $A^*$, local stability is determined by the sign of $f'(A^*)$. If $f'(A^*) < 0$, then the equilibrium is locally stable. If $f'(A^*) > 0$, then the equilibrium is unstable.

Using $x = A - A_0$, we have $f'(A^*) = -(\beta + \delta_1) - 2\delta_2 x^*$.

For the upper equilibrium $A_+^*$, this gives $f'(A_+^*) = -\sqrt{\Delta} < 0$. Therefore, the upper equilibrium is locally stable.

For the lower equilibrium $A_-^*$, this gives $f'(A_-^*) = \sqrt{\Delta} > 0$. Therefore, the lower equilibrium is unstable.

The analysis shows that the unified model exhibits a threshold structure similar to that observed in the Amrita–Imroz model. However, in the unified model, the positions of the two equilibria depend on both Sahir-related memory and Imroz-related support, represented by $S_0$ and $I_0$, respectively.

**6.5 Threshold Analysis**

The discriminant $\Delta = (\beta + \delta_1)^2 + 4\delta_2(\alpha S_0 + \gamma I_0 - \beta A_0)$ shows how the equilibrium structure changes with the parameters. Increasing $S_0$ increases the contribution of Sahir-related stimulus. Increasing $I_0$ or $\gamma$ increases the contribution of Imroz-related support. In each case, the term $\alpha S_0 + \gamma I_0$ becomes larger, and this shifts the equilibrium structure.

A threshold occurs when $\Delta = 0$.
For example, if $I_0 > 0$, a critical value of $\gamma$ is obtained from

$$(\beta + \delta_1)^2 + 4\delta_2(\alpha S_0 + \gamma I_0 - \beta A_0) = 0.$$

Solving for $\gamma$, we get

$$\gamma_c = \frac{\beta A_0 - \alpha S_0 - \dfrac{(\beta + \delta_1)^2}{4\delta_2}}{I_0}.$$

When $\Delta > 0$, two equilibria exist: a lower unstable threshold and an upper stable equilibrium. When $\Delta = 0$, the two equilibria merge, and when $\Delta < 0$, there is no real equilibrium.

This threshold should be interpreted mathematically, not biographically. It does not mean that a real emotional event occurs at a precise numerical value. It only shows how the model changes when the strengths of memory-related and companionship-related inputs are varied.

### 6.6 Numerical Simulations and Interpretation

Numerical simulations compare the unified formulation with the two preceding models and assess the effects of variations in $S(t)$ and $I(t)$ on the modelled emotional state. The simulations use $S(t) = S_0 e^{-\lambda t}$, and $I(t) = I_0$. This choice is consistent with the literary-biographical interpretation considered here. The Sahir-related influence, associated with memory, recollection, and longing rather than sustained physical presence, is represented by an exponentially decaying input. In contrast, the Imroz-related influence reflects companionship, care, and continuing support and is therefore modelled as a constant input. With these assumptions, the Sahir-related stimulus primarily affects the initial response, whereas the later dynamics are governed more strongly by the persistent Imroz-related input and the regulatory terms. The parameter values are chosen in line with the simulations presented in Sections 4 and 5 and are $A(0) = A_i = 3$, $A_0 = 5$, $S_0 = 10$, $I_0 = 4. \alpha = 1.2$, $\beta = 0.4$, $\gamma = 1, \delta_1 = 0.5$, $\delta_2 = 0.1$, $\lambda = 0.1$, $\mu = 0.05$. These values are illustrative rather than estimated from biographical data and are used only to demonstrate the model's qualitative behaviour.

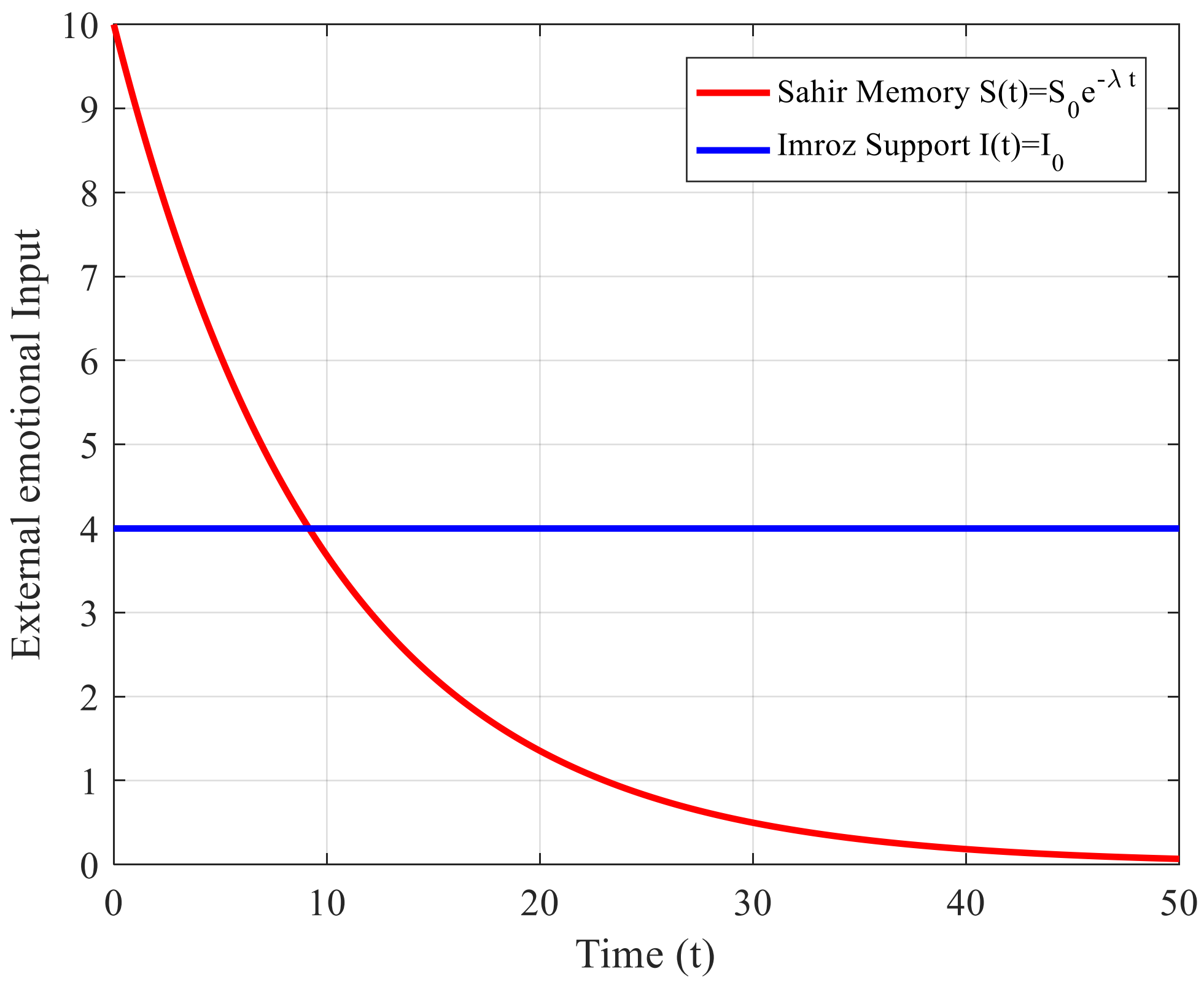

**Figure 7.** External Emotional inputs in the unified Amrita-Sahir-Imroz model

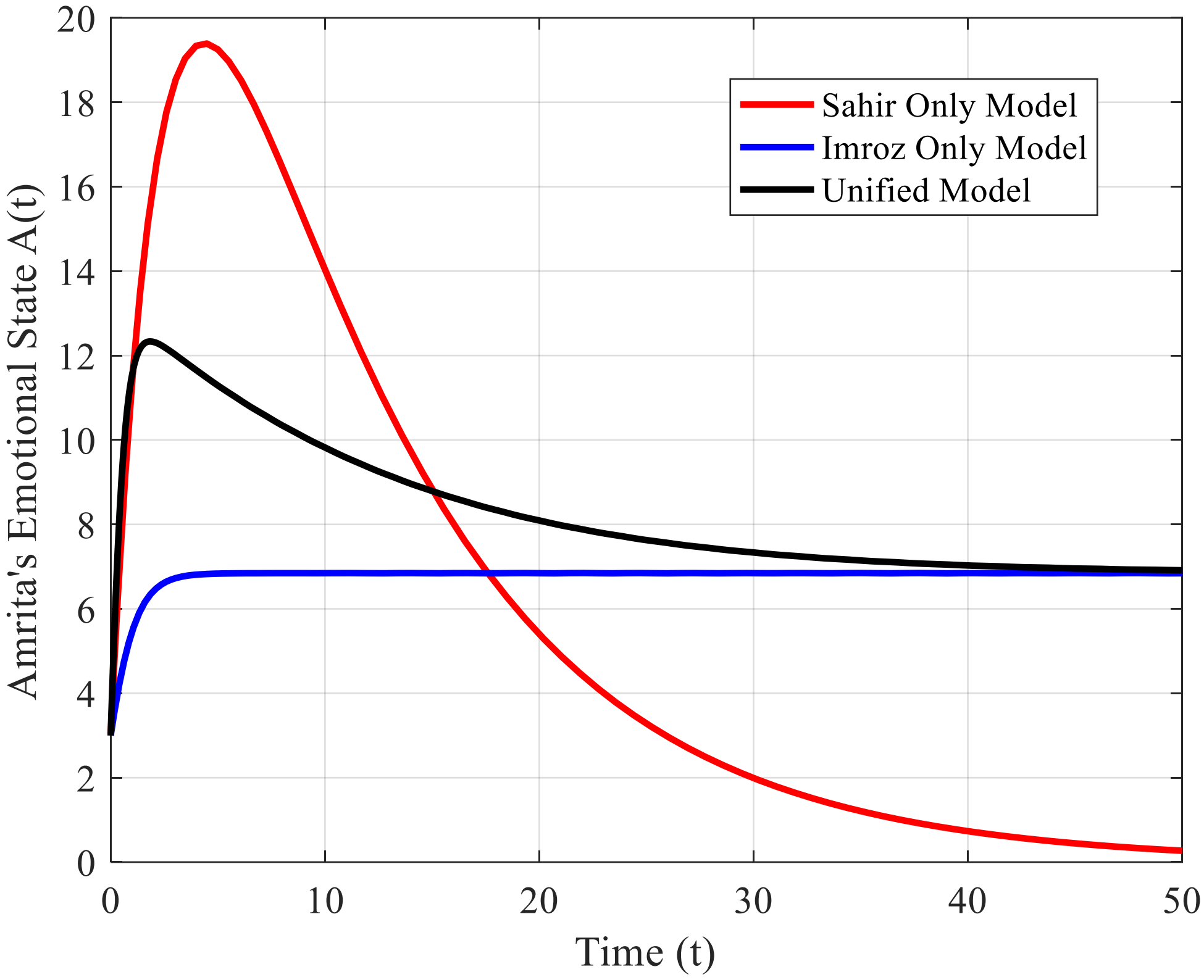

**Figure 8.** Comparison of Amrita's Emotional state $A(t)$ under Sahir-only, Imroz-only, and unified models

As described above, the two external inputs used in the unified Amrita–Sahir–Imroz model are shown in Figure 7: the Sahir-related decaying stimulus and the Imroz-related constant input. Figure 8 compares the modelled emotional state $A(t)$ for three cases: the Sahir-only model, the Imroz-only model, and the unified Amrita-Sahir-Imroz model. The Sahir-only response is modelled by excluding the Imroz-related input and regulatory terms, leaving the Sahir-related stimulus and natural fading to govern the dynamics. In the Imroz-only case, the Sahir-related stimulus is removed, and the emotional state evolves under the influence of companionship-related input and regulation towards the reference emotional level, with $\beta \neq 0$. The unified model incorporates both influences within a single-state formulation. The Sahir-only curve reflects the effect of a memory-related stimulus that gradually weakens with time. The Imroz-only curve shows the stabilizing role of companionship-related input within the admissible range of the model. In the unified case, the initial response reflects the influence of Sahir-related memory and recollection, while the sustained Imroz-related input and regulatory terms guide the modelled emotional state towards a stable range. Thus, the unified model brings together memory-driven longing and companionship-driven stabilization within a common dynamical framework.

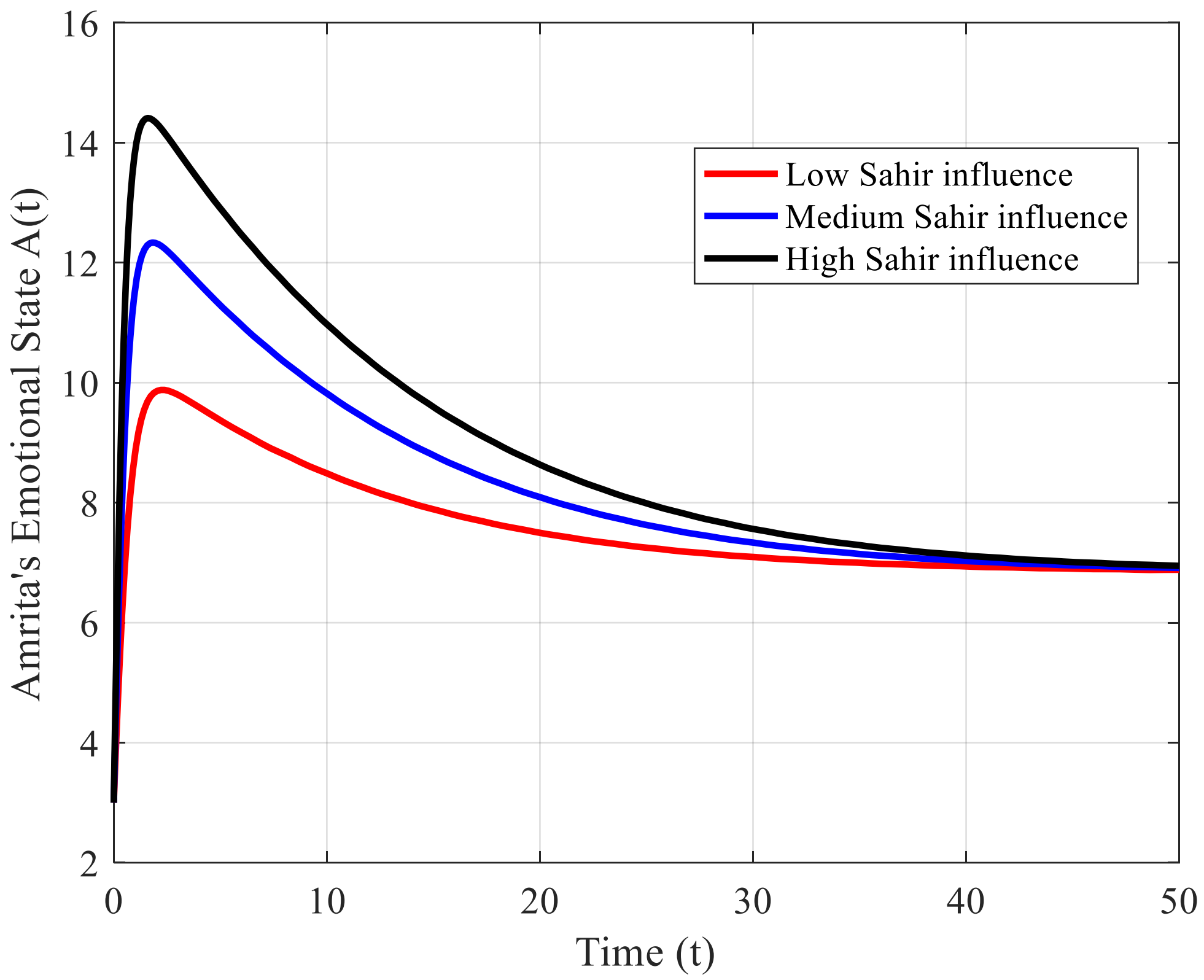

**Figure 9.** Amrita's emotional state $(t)$ in the unified model for different Sahir-related stimulus amplitudes

Figure 9 depicts Amrita's emotional state $(t)$ in the unified Amrita-Sahir-Imroz model for three different values of the initial Sahir-related stimulus amplitude. The Imroz-related input is kept the same in all cases. The three curves correspond to low, moderate, and high Sahir-related stimulus amplitudes ($S_0 = 5, 10$ and $15$). In all cases, in line with the literary-biographical interpretation, the stimulus is assumed to decay over time, reflecting the gradual fading of memory, recollection, and longing rather than an enduring presence. The curves indicate that a stronger Sahir-related stimulus produces a higher initial value of $A(t)$. This effect diminishes over time as the stimulus decays, and the influence of natural fading, Imroz-related input, and regulatory terms becomes increasingly evident. For the parameter values considered, all trajectories remain bounded and approach a regulated long-term level. Figure 10 shows the effect of changing the Imroz-related input level in the unified Amrita-Sahir-Imroz model. The Sahir-related stimulus is kept fixed as a gradually decreasing input, while the Imroz-related input is varied at weak, moderate, and strong levels ($I_0 = 2, 4$ and $6$). It allows us to observe the role of companionship-related support while keeping the memory-related component the same as a gradually decreasing. The trajectories indicate that increasing the Imroz-related input shifts the modelled emotional state toward a higher and more persistent level.

Figures 9 and 10, interpreted together, therefore suggest that the Sahir-related stimulus primarily governs the initial response, whereas the subsequent dynamics are determined more strongly by sustained Imroz-related support and regulatory effects. Unlike the Sahir-related stimulus, whose influence is concentrated primarily in the initial phase, the Imroz-related input has a stronger effect on long-term behaviour. As the Sahir-related stimulus declines, the sustained Imroz-related input, together with the regulatory terms, drives the state toward a stable range.

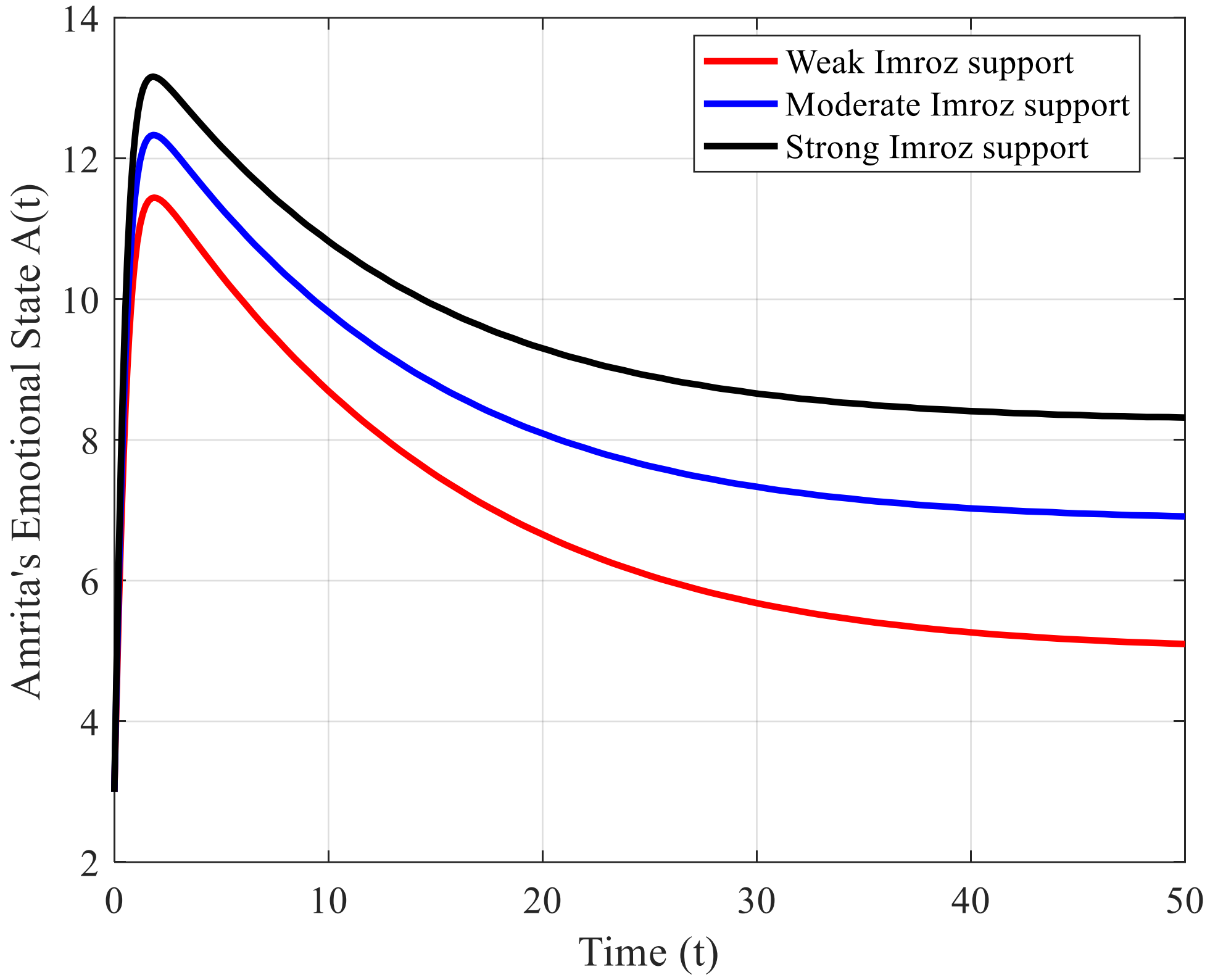

**Figure 10.** Amrita's emotional state $(t)$ in the unified model for different Imroz-related input levels

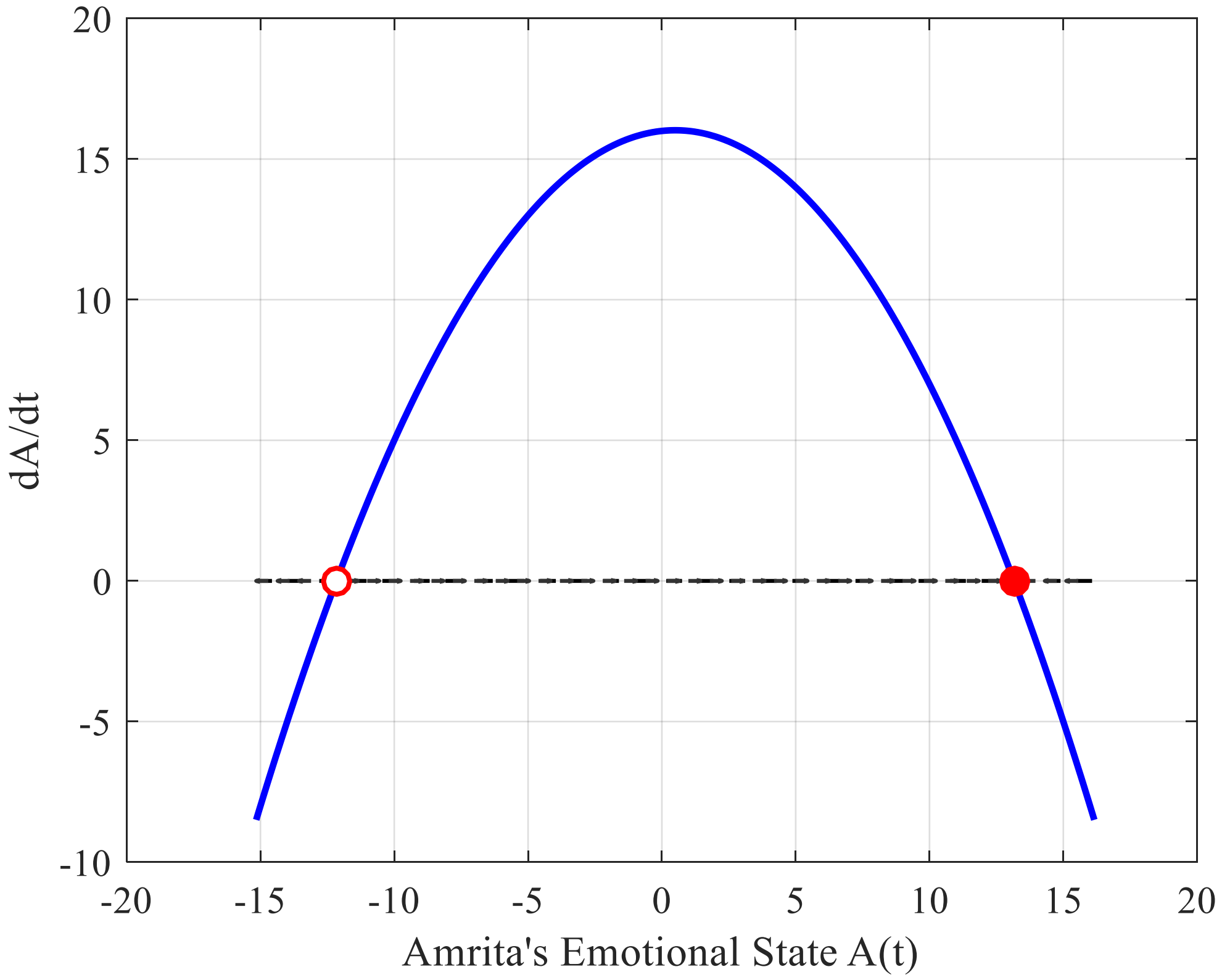

**Figure 11.** Phase-line diagram of the unified Amrita—Sahir-Imroz model considering constant Sahir-related and Imroz-related input, i.e., $S(t) = S_0$ and $I(t) = I_0$.

Figure 11 presents the phase-line diagram of the unified Amrita-Sahir-Imroz model for constant Sahir-related and Imroz-related inputs. The curve shows the right-hand-side function $f(A)$ plotted against the modelled emotional state $A$. The points where the curve crosses the horizontal axis give the equilibrium values of the system. The sign of $f(A)$ gives the direction of change of the modelled emotional state. Positive values of $f(A)$ correspond to an increase in $A(t)$, while negative values correspond to a decrease. The right (larger) equilibrium point is locally stable, since nearby trajectories are directed toward it and the left (smaller) equilibrium point marks an unstable threshold. This agrees with the analytical stability result derived above. For the selected parameter values and admissible initial conditions, the emotional state $A(t)$ in the unified model moves toward the stable upper equilibrium. Figure 12 presents the equilibrium values of the unified model as the Sahir-related stimulus amplitude $S_0$ is varied. The Imroz-related input level and all other parameters are kept fixed. The upper branch gives the locally stable equilibrium, while the lower branch gives the unstable threshold. As $S_0$increases, the stable equilibrium value $A^*$ also increases, which indicates that a stronger Sahir-related stimulus, associated with memory, recollection, and longing, shifts the modelled emotional state toward a higher equilibrium level. The lower branch should be understood as a threshold and the initial values below it do not belong to the stabilizing range considered here. The figure therefore shows that the long-term behaviour of the unified model is sensitive to the Sahir-related stimulus amplitude. Combining this observation with Figure 9, we may infer that the Sahir-related component affects not only the early response of $A(t)$, but also the equilibrium values obtained under constant-input conditions.

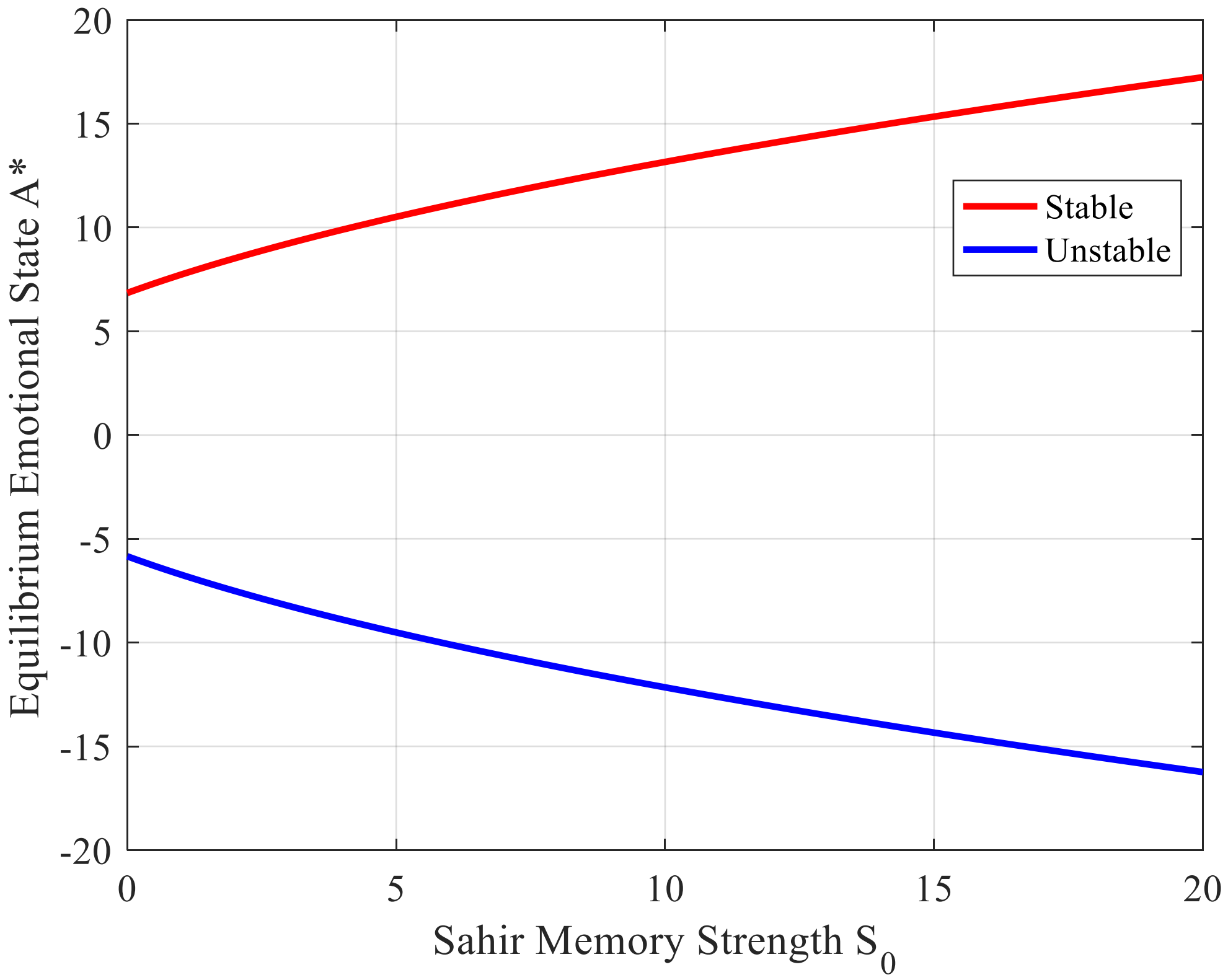

**Figure 12.** Equilibrium sensitivity to Sahir-related stimulus amplitude

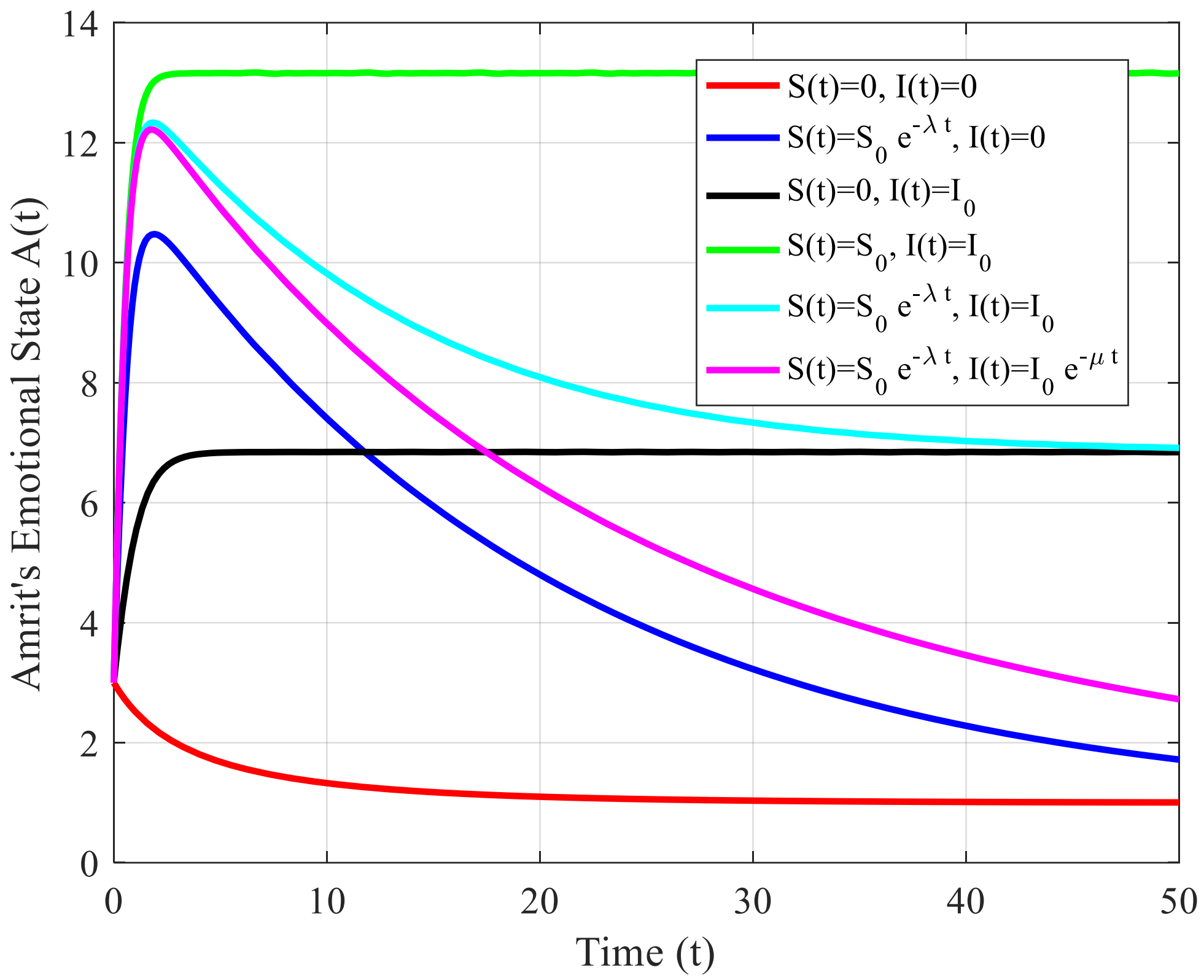

**Figure 13.** Six representative numerical cases of the unified Amrita-Sahir-Imroz model

Figure 13 compares six representative numerical cases of the unified Amrita-Sahir-Imroz model. Case I considers intrinsic regulation only, where both the Sahir-related stimulus and the Imroz-related input are absent. Case II gives the Sahir-only response, in which the modelled emotional state is influenced by memory-related stimulus and natural fading. Case III gives the Imroz-only response, where the Sahir-related stimulus is absent and the motion is governed by companionship-related input and regulation around the reference emotional level. Case IV considers constant Sahir-related and Imroz-related inputs together. Case V uses a gradually weakening Sahir-related stimulus with constant Imroz-related input. Case VI allows both inputs to decrease with time. The figure illustrates how the unified single-state model accommodates different input configurations within a common mathematical framework. In the context of the present literary-biographical analysis, the case involving a decaying Sahir-related stimulus and a constant Imroz-related input is particularly relevant. It reflects the gradual fading of memories and recollections over time, while companionship, care, and continued emotional support remain influential in determining the subsequent evolution of $A(t)$

## 7. Concluding Discussion, Limitations and Future Scope

The three models developed in this study provide a progressive mathematical description of selected emotional patterns associated with Amrita Pritam's literary-biographical context. Model I represents the Sahir-related component through memory, recollection, reinforcement, and natural fading. It shows that the modelled emotional state may decay in the absence of stimulus, approach a sustained level under constant stimulus, or gradually weaken when the stimulus itself decreases with time. This makes the model suitable for representing a memory-related emotional response that persists for some time without being treated as permanent. Model II shifts the focus from memory-driven persistence to companionship-driven stabilization. The Imroz-related input is represented together with regulation around a reference emotional level. Under constant input, the model gives a lower unstable threshold and an upper locally stable equilibrium. This indicates that companionship-related support can move the modelled emotional state toward a stable level, provided the initial condition lies within the admissible range of the model. Model III combines the Sahir-related stimulus with the Imroz-related input to form a unified single-state model. In this model, memory-related longing, natural fading, companionship-related support, and nonlinear regulation jointly influence the same modelled emotional state. The equilibrium analysis shows that the long-term behaviour depends on both the Sahir-related stimulus amplitude and the Imroz-related input level. Thus, the unified model provides a simple mathematical way to represent how memory-driven and companionship-driven influences may coexist within one emotional trajectory.

The study has certain limitations. The single variable, used as a symbolic modelled emotional state of Amrita Pritam, should not be understood as a direct measurement of Amrita Pritam's private emotion. The parameter values used in the numerical simulations are illustrative and are chosen to show the qualitative behaviour of the models rather than to estimate biographical or psychological quantities. A single scalar variable also cannot represent the full complexity of Amrita Pritam's literary-biographical experience, which includes memory, love, loss, partition, creativity, social context, and personal choice. The Sahir-related and Imroz-related terms should therefore be understood only as model-based representations of two emotional modes: memory-driven longing and companionship-driven stabilization.

Future work may extend this scalar formulation to a multivariable model. Separate state variables may be introduced for longing, companionship, creative expression, and historical or biographical forcing. Such a model would allow different emotional dimensions to interact more explicitly. Delayed or memory-dependent terms may also be considered to represent the persistence of recollection and the delayed effect of emotional or creative experience. In addition, coupled formulations could be developed to study interactions among separate emotional components rather than representing them through a single state variable. The present paper may therefore be viewed as a first mathematical formulation, preparing the ground for a richer multidimensional framework.

Overall, the study shows that a single-state dynamical framework can capture several useful qualitative features, including fading, persistence, equilibrium, stabilization, response to external inputs, and threshold behaviour. The proposed models are not intended to replace literary interpretation. Instead, they provide a mathematical framework for describing and exploring selected emotional patterns within a literary-biographical context.

## Author Contributions

VP: Conceptualization, Formal Analysis, Investigation, Methodology, Validation, Visualization, Writing–original draft, Writing–review and editing. TS: Conceptualization, Formal Analysis, Investigation, Methodology, Validation, Visualization, Writing–original draft, Writing–review and editing.

## Declaration of generative AI and AI-assisted technologies

During the preparation of this manuscript, the authors used ChatGPT (OpenAI) to improve sentence structure, clarity, and readability. The authors reviewed and revised the suggested text as appropriate and take full responsibility for the content of the article.